\documentclass{IEEEtran4PSCC}
\ifCLASSINFOpdf
   \usepackage[pdftex]{graphicx}
\else
   \usepackage[dvips]{graphicx}
\fi
\usepackage{xcolor}
\usepackage{hyperref}
\usepackage{xspace,tikz,array}
\usepackage{multirow}
\usepackage{bm}
\usetikzlibrary{arrows,automata,shapes,positioning}
\usepackage{scrextend}
\usepackage{graphicx}

\usepackage[cmex10]{amsmath}
\ifCLASSOPTIONcompsoc
  \usepackage[caption=false,font=normalsize,labelfont=sf,textfont=sf]{subfig}
\else
  \usepackage[caption=false,font=footnotesize]{subfig}
\fi

\usepackage{booktabs}
\usepackage{calc}

\renewcommand{\aa}{\bm{a}}
\newcommand{\xx}{\bm{x}}
\newcommand{\yy}{\bm{y}}

\renewcommand{\ss}{\bm{s}}
\newcommand{\ee}{\bm{\epsilon}}
\renewcommand{\tt}{\bm{\tau}}

\makeatletter
\let\old@ps@headings\ps@headings
\let\old@ps@IEEEtitlepagestyle\ps@IEEEtitlepagestyle
\def\psccfooter#1{%
    \def\ps@headings{%
        \old@ps@headings%
        \def\@oddfoot{\strut\hfill#1\hfill\strut}%
        \def\@evenfoot{\strut\hfill#1\hfill\strut}%
    }%
    \def\ps@IEEEtitlepagestyle{%
        \old@ps@IEEEtitlepagestyle%
        \def\@oddfoot{\strut\hfill#1\hfill\strut}%
        \def\@evenfoot{\strut\hfill#1\hfill\strut}%
    }%
    \ps@headings%
}
\makeatother

\psccfooter{%
        \parbox{\textwidth}{\hrulefill \\ \small{21st Power Systems Computation Conference} \hfill \begin{minipage}{0.2\textwidth}\centering \vspace*{4pt} \includegraphics[scale=0.06]{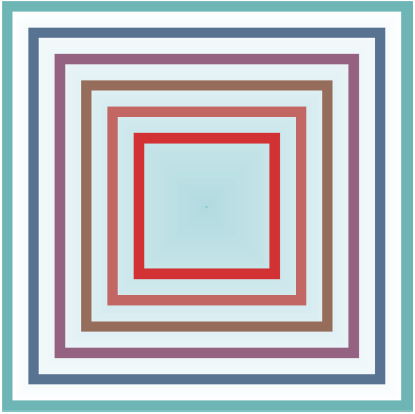}\\\small{PSCC 2020} \end{minipage} \hfill \small{Porto, Portugal --- June 29 -- July 3, 2020}}%
}

\begin{document}
%
\title{Learning to run a power network challenge \\ for training topology controllers}

\author{
\IEEEauthorblockN{Antoine Marot, Benjamin Donnot\\ Camilo Romero, Balthazar Donon}
\IEEEauthorblockA{
R\'eseau de Transport d'\'Electricit\'e \\
Paris, France}
\and
\IEEEauthorblockN{Marvin Lerousseau }
\IEEEauthorblockA{
INSERM and CVN \\ CentraleSupélec and INRIA \\ Paris, France}
\and
\IEEEauthorblockN{Luca Veyrin-Forrer, Isabelle Guyon}
\IEEEauthorblockA{TAU group of Lab. de Res. en Informatique \\
UPSud/INRIA Universi\'e Paris-Saclay \\
Paris, France }
}


\maketitle

\begin{abstract}
    For power grid operations, a large body of research focuses on using generation redispatching, load shedding or demand side management flexibilities. However, a less costly and potentially more flexible option would be grid topology reconfiguration, as already partially exploited by Coreso (European RSC) and RTE (French TSO) operations. Beyond previous work on branch switching, bus reconfigurations are a broader class of action and could provide some substantial benefits to route electricity and optimize the grid capacity to keep it within safety margins. Because of its non-linear and combinatorial nature, no existing optimal power flow solver can yet tackle this problem. We here propose a new framework to learn topology controllers through imitation and reinforcement learning. We present the design and the results of the first ``Learning to Run a Power Network'' challenge released with this framework. We finally develop a method providing performance upper-bounds (oracle), which highlights remaining unsolved challenges and suggests future directions of improvement.
\end{abstract}


\section{Introduction}

    Grid operators are in charge of ensuring that a reliable supply of electricity is provided everywhere, at all times. However, their task is becoming increasingly difficult under the current steep energy transition. On one hand, we observe the advent of intermittent renewable energies on the production side and of prosumers on the demand side, coupled to the globalization of energy markets over a more and more interconnected European grid. This brings a whole new set of actors to the power system, adding lots of uncertainties to it. On the other hand, recent improvements in terms of energy efficiency have put an end to the total consumption growth, and \textit{de facto} to the growth in revenues, thus limiting new costly investments. Public acceptance with regards to the installation of new infrastructure is also a growing issue. This shifts the way we traditionally develop the grid, from expanding its capacity by building new power lines, to optimizing the existing one as it is, closer to its limits, with rather digital means and every flexibilities at our disposal. 

    Currently, operators must analyze massive amounts of data to take complex coordinated decisions over time with higher reactivity, to operate under an ever more constrained environment with greater control. An under-exploited flexibility, which could alleviate in part the problem, is the {\bf grid topology}, as already pointed out by early work \cite{TopologyOld}. This avenue is explored in the present paper. As of today, it is still beyond the state-of-the-art to control optimally such grid topology ``at scale'', beyond the level of ``branch switching'' \cite{fish2008BrSwitching} (e.g. considering more complex actions such as ``bus splitting''), because of the non-linear and combinatorial nature of the problem. Lately, we demonstrated an augmented expert system ability \cite{ExpertSystem} to discover steady-state tactical solutions to unsecure grid states by relying on bus splitting. Its acceptable computational time opens new perspectives for a revival of such class of actions. In addition, novel and more flexible actions should also be considered today, intervening not only instantaneously as a {\em tactic} rather independently of other decisions, but over a time horizon as a {\em strategy} to manage more numerous overlapping and interfering decisions. 
    To exploit such a complex array of actions, optimal control methods [5] have been explored but are hard to deploy in control rooms because of limited computing budget constraints.

    With the latest breakthroughs in Artificial Intelligence (AI) from AlphaGo at Go \cite{AlphaGo} and Libratus at Poker \cite{Libratus}, Deep Reinforcement Learning (RL) seems a promising avenue to develop a control algorithm, \textit{a.k.a.} an artificial agent, able to operate a complex power system at scale near real-time and over time, assisted by existing advanced physical grid simulators. Recent work \cite{donnot-NeuralNet} has already shown the value of Deep Learning for accelerating power flow computation and risk assessment applications for power systems, demonstrating its ability to model such system behavior, adding more credit to the Deep RL potential. RL formulations have already been applied to specific power system related problems (see \cite{RLreview} for a review), but not for continuous power system real-time operations, a problem for which no test cases existed today, probably limiting any subsequent development. Related applications concern the unit commitment problem with an interesting multi-stage formulation \cite{GalDalal}, but on an infra-day timeframe with rather simplistic assessment of intra-day real time operation. These authors also promote leveraging the capabilities of Deep RL as a catalyst for successful future works. They finally insist on the real-world challenge of Safe-RL to manage for instance a power system for which contingencies, and related overloads, rarely occur but need to be closely managed, to avoid cascading failures and subsequent blackouts. More broadly, Safe-RL is a hot research topic in the whole RL community to eventually address real-world challenges for critical systems \cite{RLchallenges}.

    In line with those recent development and in order to foster further advances both within the power system and RL communities, we open-sourced a new platform to build and run power system synthetic environments to further develop and benchmark new controllers for continuous near-real time operations. We indeed built and released a first IEEE14 environment test case upon which we organized a competition ”Learning to run a Power Network” (L2RPN) with an emphasis on the challenging use of topological flexibilities and the safety robustness requirement. The L2RPN competition which we will present and analyze here, takes some inspiration from the « Learning to run » \cite{L2R} competition, whose goal was to learn a controller of a human body to walk and run. This was an opportunity for bio mechanical researchers to successfully address their problem together with the RL community.

    The following paper is organized as follows: first we present an overview of the objective and results of this first L2RPN competition. We further review the design and modelling of the related test environment. We then propose a conceptual Markov Decision Process framework to analyze the nature of the problem for a given test environment. Finally we describe the results through a comparative analysis of the best submitted agents and other baselines, and give conclusions.

\section{The L2RPN competition}

    \paragraph{Challenge overview}
    
\begin{figure}[h]
            \subfloat[Leaderboard]{
            \includegraphics[width=5cm, height=3.2cm]
            {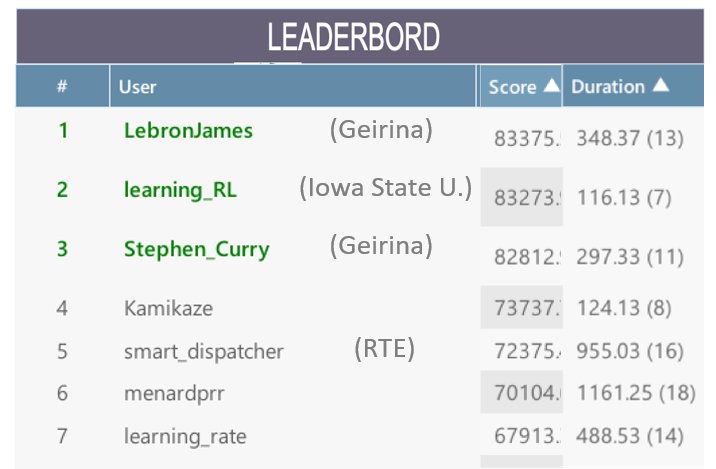}
            \label{fig:Leaderboard}
            }
            \subfloat[Score  function f(Margin$_l$)]{
            \includegraphics[width=3.5cm, height=2.8cm]
            {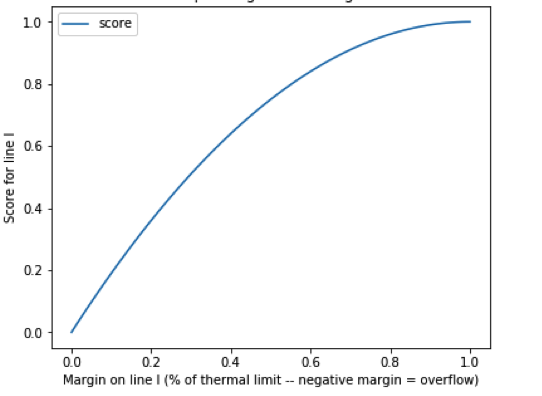}
            \label{fig:Scores}
            }
            \caption{ \ref{fig:Leaderboard} Final leaderboard of the L2RPN competition with cumulated scores and computation time. \ref{fig:Scores} Instantaneous line margin score function }
        \end{figure}

        This first L2RPN competition ran over 6 weeks starting on May 2019 over the Codalab challenge platform. It was based on the pypownet \cite{Pypownet} open-sourced platform\footnote{\label{website}GitHub for pypownet framework: \url{https://github.com/MarvinLer/pypownet}} relying on openGym RL framework. 100 participants signed in, 15 of which were particularly active with many submissions every week. Figure \ref{fig:Leaderboard} showed the 7 best participants who all achieved interesting and successful scores, with a mix of RL, Machine Learning (ML), expert system and tree search approaches we comment in section \ref{sec:res}. They all succeeded in beating simple baselines used to design the challenge, and RL approaches achieved the best results.


        The goal of the competition was to operate the power grid in real-time over several days at a 5 minute time-resolution. More precisely, the aim was to manage the powerflows \textbf{i} (in Amperes) at every time-step \textbf{t}, given injections \textbf{x}, \textit{aka} production and loads, and grid topology denoted by ${\bm{\tau}}$. The score function gave incentives to optimize the margins through all lines, in order to maximize the overall residual grid capacity \textbf{$\Delta$Cg}, given power lines capacities \textbf{imax$_{\text{l}}$}, \textit{aka} thermal limits:
        \begin{equation}
            Margin_{l}(t)=\max\left(0,1-\frac{i_l(t)}{imax_l}\right), \ \forall l \in L
        \end{equation}

        To manage the grid, an agent could only use topological actions in \textbf{A$_{\tau}$} on lines \textbf{L} and substations \textbf{Sub} in that challenge, while meeting further operational constraints on actions and overloads. Some actions could be illegal, violating some operational rules further described in Section \ref{sec:design}, or result in a diverging powerflow computation, most likely due to some voltage issues. Choosing an illegal action resulted in a null score at that time-step. To assess an agent performance, the following score at time-step \textbf{t} was used as a proxy for \textbf{$\Delta$Cg} :
        
        \begin{equation}
        Score(t) =
        \begin{cases}
            0 & \text{ if illegal action or divergence} \\
            \Delta Cg(t) &=\sum\limits_{l=1}^{n_l} f(Margin_{l}(t)) 
        \end{cases}
        \end{equation}

        Every line somehow contribute to the real-time power grid capacity, represented by the sum in the score. But already loaded lines should affect more the residual capacity of the grid as they soon become bottlenecks to transmit more flows. This is why more value should be given on increasing the margins of already loaded lines, while increasing the margin of a line not very loaded does not add up to the overall capacity. We chose $\bm{f(x)}=1-(1-x)^2$ shown in Figure \ref{fig:Scores} in that prospect. Optimizing \textbf{$\Delta$Cg} eventually contributes to minimizing power losses on the grid, smoothing the power line usage to avoid materials aging too quickly, and maximizing the grid flexibility which are all today's challenges for power grids in most developed countries.

        Finally, and most importantly, overload \textbf{Ov$_{\text{l}}$} (that happen when $i_{l} >= imax_{l}$) had to be solved dynamically to avoid line disconnection and further cascading failures. Any cascading failure resulted in a game over and a null score for a scenario \textbf{s}, putting a strong penalty on not being secure:
        \begin{equation}
        Score(s)=
        \begin{cases}
            0, & \text{if game-over  during  s} \\
            \sum\limits_{t=0}^T Score(t), & \text{otherwise}
        \end{cases}
        \end{equation}
        Formalizing the score objective as a maximization problem of positive gains allows us to easily define a penalty as a null score if some rules of the environment are violated. 

    \paragraph{Evaluation}

        Along the competition, participants could submit their agent on the Codalab platform\footnote{The challenge has been re-enacted and is open for post-challenge submissions at \url{https://competitions.codalab.org/competitions/20767}} to be tested on unseen validation scenarios. A score was computed for each scenario and revealed to the participants, in addition to the time-step of game over if one occurred. At the end of the competition, participants were assessed on new 10 secret test scenarios similarly chosen, as described in Section \ref{sec:res}, with no other information than their cumulative score. Finally, to reflect one main characteristic of real-time operations for which time is limited, resulting in a trade-off between exploration and exploitation, participants would only get a score if they managed to finish their scenario within an allocated 20-minute time budget as shown on Figure \ref{fig:Leaderboard}. It was calibrated as being 10 times the running time of the most simple baseline: a do-nothing (DN) agent. Expert Systems and tree-search approaches appeared to be slow, since they relied more extensively on simulations, and had to limit their exploration to meet the time constraint.

    \paragraph{Benefits of a Challenge}

        While reproducibility is a hot issue for scientific research and further model deployment, a challenge format looks appealing to avoid some pitfalls. Indeed, a challenge acts as a benchmark whose goal is to decouple the problem modeling from a solver implementation. 
        Compared to the fine tuning of ones own  method, a challenge aims at giving equal chance to every field expert to tune its solution in a given time slot and compare against one another. Its success also depends on the clarity of the submitted problem, on the ergonomy and robustness of its platform, on the transparency of its results, with an intrinsic need of reproducibility to faithfully deliver ones score and reliably announce the winners. Finally, winners had to open-source their code to obtain the price, further enforcing reproducibility. 

\section{Challenge Design} \label{sec:design}

    The difficulty of setting up this kind of challenge lies in the fact that one has to design a whole synthetic and interactive environment to test controllers (like a video game), and not just a fixed dataset. We designed it following 3 guiding principles: 

    \begin{enumerate}
        \item	Realism : the environment should represent real-world power system operational constraints and distributions.
        \item	Feasibility : solutions should exist to finish scenarios under the available actions given game over conditions.  
        \item	Interest : harder scenarios should be challenging to finish and most scenarios should be complex to optimize.
    \end{enumerate}

    There is a natural trade-off in making a game interesting and feasible: a most interesting game will more likely be more difficult with potential feasibility issues arising. 

    Now a Grid World, \textit{aka} an environment for a power grid, is the combination of the following components:
    
    \begin{itemize}
        \item	A power grid (with substations and powerlines of different characteristics) and a grid topology $\bm{\tau}$
        \item	Real-time injections \textbf{x(t)}, next time-step forecasts \textbf{\^{x(t+1)}}
        \item	Lines Capacities \textbf{imax$_{\text{l}}$} and additional	operational rules 
        \item	Events such as maintenance and contingencies 
    \end{itemize}

    The power flows $i_l(t)$ are computed by a power-flow simulator.The platform then allows some interactions for an agent with that environment through actions $a_{\tau}(t)$. Participants had the option to simulate the effect of their action before choosing one, but using some computation time budget. To assess how feasible and interesting an environment is for agents, we defined the following simple baseline agents :
    
    \begin{itemize}
        \item a ``do nothing'' (\textcolor{blue}{DN}) agent that remains in the reference topology $\tau_{ref}$ and which is already robust most of the time, much better actually than taking random, and mostly stupid, actions. 
        \item a single topology agent (\textcolor{blue}{DN$_{\tau}$}), which is a DN agent running in a constant topology $\tau$ distinct from $\tau_{ref}$.
        \item a greedy (\textcolor{blue}{GR}) agent that simulate do-nothing and all unitary actions at a given time-step, and take the most immediately beneficial one.
    \end{itemize}
    
    These agents can finish all together most scenarios, while highlighting hard to complete ones.
    Let's now describe in more details the different steps of the challenge design.

    \paragraph{Choosing a Power Grid and define topology $\tau$} 

        To be realistic, we first chose a grid among common IEEE power grids. For that first challenge, we chose a minimalist grid to better ensure feasibility and help analysis, but yet a grid for which topological actions could still be useful to manage it. Such a grid needed to be meshed with several electrical paths. We hence chose the IEEE14 grid as it is a meshed grid with 2 West and East corridors between a meshed South Transmission Grid and a meshed North Distribution Grid. Even with only 14 substations, the number of potential overall configurations, hence the combination of actions, is tremendous. For such a size, we however did not consider the occurrence of contingencies and maintenance since it will most often lead to infeasibilities. Overloads were only the result of peak productions or peak loads in certain areas of the grid. 

        The reference topology $\tau_{ref}$ is the base case topology, fully meshed, with every lines in service and single electrical node per substations. However, up to two electrical nodes are possible per substation, modeled as a 2-bus bar. 
        
        The topology can be changed by actions $a_{\tau}(t) \in A_{\tau}$, that can be of two kinds :
        
        \begin{itemize}
            \item	Binary Line switching $a_l$ ($2^{20}$ possible actions) 
            \item	Bus bar switching of elements at a substation $a_{bus_i}$ – ($2^{20*2+16}$) for 20 lines and 16 injections
        \end{itemize}
        
There are as many actions as there are topology configurations.
Eventually, the number of reachable configurations at a given
time is limited by considering some operational rules.

        \begin{figure}[h]
            \hspace*{\fill}%
            \begin{center}
            \subfloat{\includegraphics[width=4.8cm, height=3cm]
            {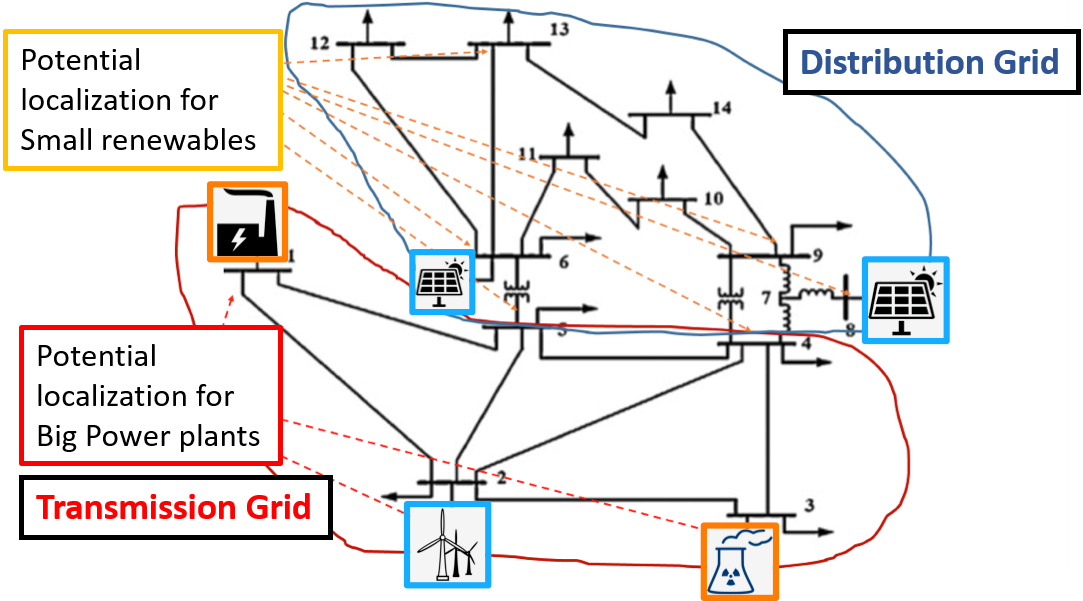}
            }
            \subfloat{\includegraphics[width=3.8cm, height=3cm]
            {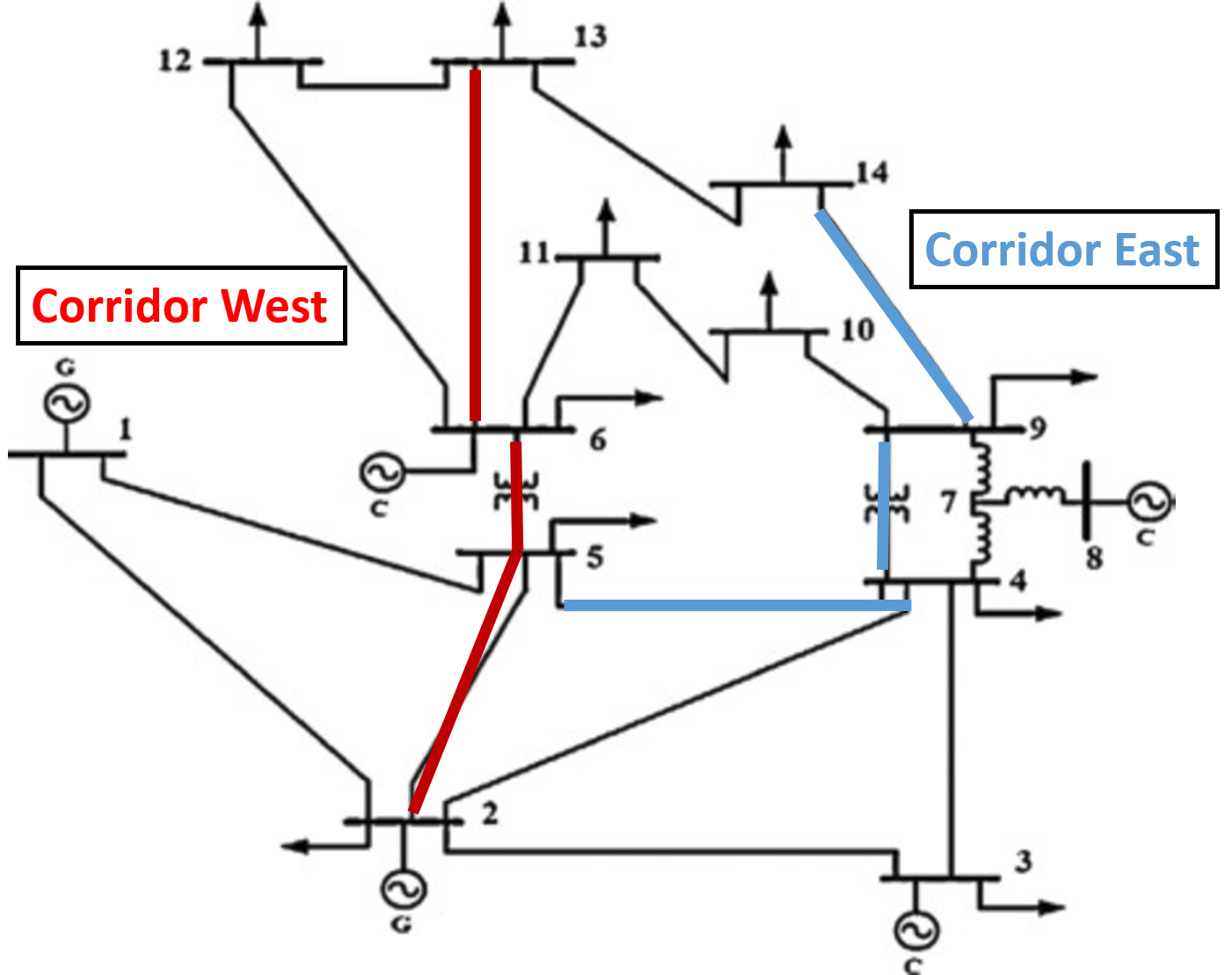}
            }
            \end{center}
            \caption{a) IEEE14 Grid and production localization. b) Two existing electrical corridors between Transmission and Distribution Grids.}
            \label{fig:IEEE}
        \end{figure}

    \paragraph{Generating Injections x(t) and Forecasts \^x(t+1)}
    
\begin{figure}[h]
            \hspace*{\fill}%
            \begin{center}
            \subfloat{\includegraphics[width=4.6cm, height=2.6cm]
            {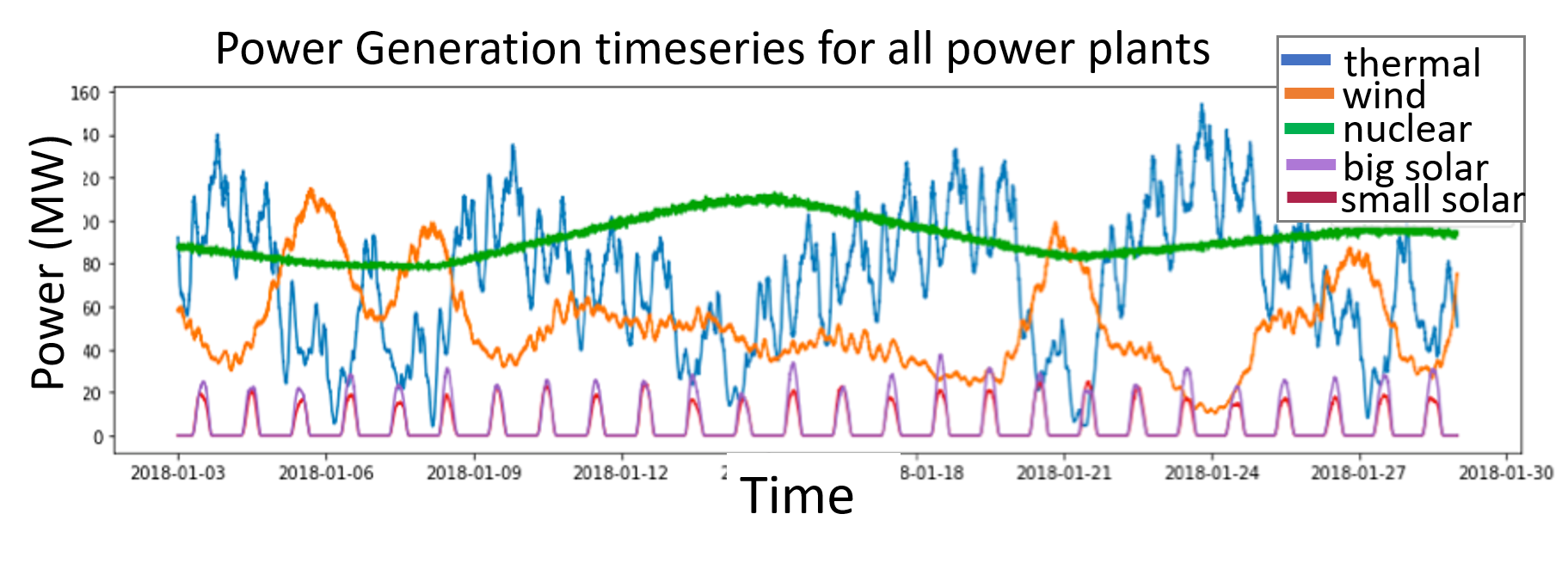}
            }
            \subfloat{\includegraphics[width=4cm, height=2.6cm]
            {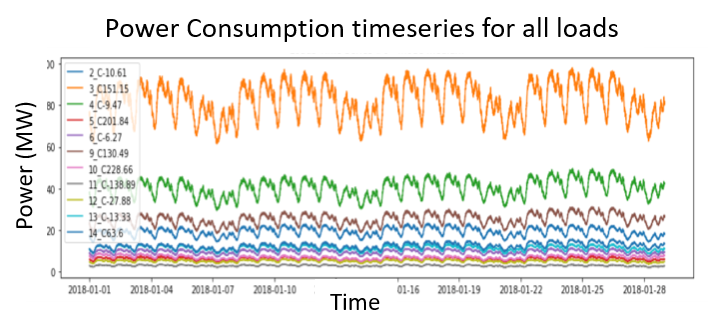}
            }
            \end{center}
            \caption{ a) Production, thermal and renewable, profiles over a typical month of January. b) Load profiles, all similar  modulo a proportional factor. }
            \label{fig:injections}
        \end{figure}

        While the IEEE 14 grid only had initially thermal production, we introduce renewable wind and solar plants as depicted on Figure \ref{fig:IEEE} to better represents today's energy mix and dynamics to consider managing resulting issues. 
        The total load consumption profile follows the French consumption one and individual loads are computed given a constant key factor from the original IEEE case on Figure \ref{fig:injections}. Solar and wind power also follow French related distribution with correlations between wind and solar.
        The nuclear power plant is a baseload, slowly varying and the thermal power plant compensates for the remaining production required for balancing (see Figure \ref{fig:injections}).
        For this challenge, we restricted  the distribution of injections to be representative of winter months over which we observe peak loads. Next time frame forecast where also provided for injections $x$ with 5\% gaussian noise uncertainty.

    \paragraph{Line capacities imax and operational rules}:
    
        \begin{figure}[!htbp]
            \begin{center}
            \includegraphics[width=6cm, height=2.5cm]
            {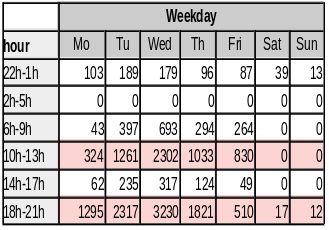}
            \end{center}
            \caption{Number of overloads per day and hours over 50 monthly representative scenarios. Overloads mainly appear on weekdays at peak load hours or under peak solar production around noon on the distribution grid.}
            \label{fig:overloads}
        \end{figure}
  
        The second part of the game design was about setting up the right thermal limits (no thermal limits are given for the IEEE14 case) so that some overloads appear, not all being easily solvable, but many of them that can be solved by at least a baseline to remain feasible. From the grid structure perspective, only 2 electrical corridors exists as illustrated on Figure \ref{fig:IEEE}. We cannot allow both of them being overloaded at the same time, since there will not exist any more path to reroute electricity with topology to relieve all overloads. We hence chose to preferably constrain line capacities on the West Corridor (buses $2 \to 5 \to 6 \to 13$), while keeping spatial consistency for thermal limits overall from a grid development perspective. Overloads eventually appear 3\% of the time over those lines in the scenarios running DN agent. Other lines had their capacities rated above the max power flow observe in the reference topology.
        
        To make the game more interesting, we also want to have a spread distribution on the time of occurrence of overloads as shown on Figure \ref{fig:overloads}.
        Running an ensemble of expertly selected $DN_{\tau}$ baselines (see Table \ref{table:tables}) solves $85\%$ of the overloads. This demonstrates that our game is both feasible, and has some difficult interesting situations (following our guiding principles). Also, the GR baseline does not perform well on all scenarios : this agent tends to get stuck in bad local configurations.

        Finally, real-world problems get complex because of operational constraints.
        We choose to model the following ones:
        \begin{itemize}
            \item reaction time -- time to react to an overload before the line get disconnected by protections, 2 time-steps here. 
            \item activation time -- there is a maximum number of actions that a human or a technology can performed in a given time period, one action per substation here. 
            \item recovery time (cooldown) -- due to physical properties of the assets, there is some time after activation before a flexibility can be reactivated, set to 3 time-steps here.
        \end{itemize}

        You can look at flexibilities as a kind of budget. When you use one, you consume part of your budget before recovering it some time after. This induces some credit assignment problem. Let's now formalize the whole problem under those settings with the generic framework of Markov Decision Processes.

        

    \section{Problem Formalization}

        Markov Decision Processes (MDPs) \cite{SuttonRL} 
        are useful and general abstractions to solve a number of problems in optimization and control. We decide to further formalize our problem here using this framework, which will help determine the nature of the problem under different settings. An MDP is generally represented by a tuple $\langle S, A, P, r,  \gamma \rangle$ with:
        
        \begin{itemize}
            \item $S$ the state space of observations from the environment.
            \item $A$ the action space, the potential agent interactions with the environment. 
            \item $P$ the stochastic transition function $p(s(t+1)|s(t),a(t))$ which computes the system dynamics. It defines a Markovian assumption. 
            \item $r(t)=r(s(t),a(t))$ the immediate reward function.
        \end{itemize}

        The {\em policy} determines the next action $\aa(t+1)$ as a function of $\ss(t)$ and $r(t)$. The policy can adapt itself (by reinforcement learning) to maximize the expectation (over all possible trajectories) of this function:
        
        \begin{equation}
            G(t) = \sum_{k=t+1}^T \gamma^{k-t-1} r(k)
            \label{eq:return}
        \end{equation}
        where $\gamma \in [0, 1]$ is the discount rate.

\begin{figure}[h]
            \hspace*{\fill}%
            \begin{center}
            \includegraphics[width=5.3cm, height=4cm]
            {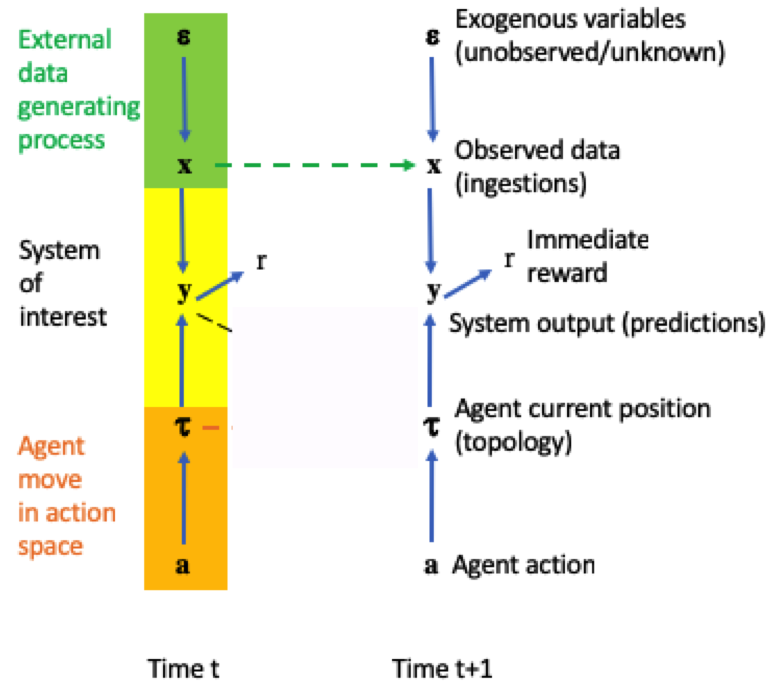}
            \label{fig:Bandits}
            \includegraphics[width=3.0cm, height=4cm]
            {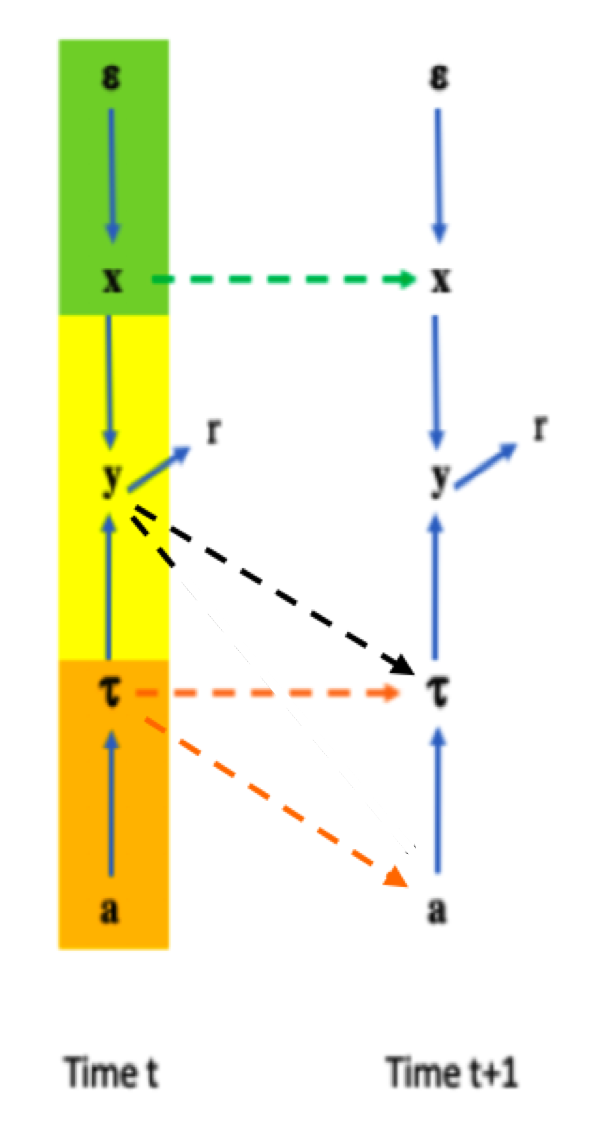}
            \label{fig:L2RPN_MDP}
            \end{center}
            \caption{\textbf{a)} Contextual Bandit framework. \textbf{b)} L2RPN MDP formalization. }
            \label{fig:MDP}
        \end{figure}

        Our L2RPN problem is actually a two-factor MDP (Figure \ref{fig:MDP}), a special case of MDP in which the state $S$ is a quadruplet:
        
        \begin{itemize}
            \item $\ee$: unobserved influences on inputs $\xx$.
            \item $\xx$: observed inputs of the system, not influenced by the actions of the agent.
            \item $\tt$: other observed inputs of the system, influenced by the actions of the agent.
            \item $\yy$: observed outputs of the system, $\yy = F(\xx, \tt)$.
        \end{itemize}
        
        Function $F$ specifies the system of interest and y is essentially here powerflows $i_l$ and overloads $Ov_l$. F is a function of {\bf two factors} $\xx$ and $\tt$ (injections and topology in our case). The `observed inputs' $\xx$ may or may not be organized in a time series. In our case they are. Injections are continuous time-series. For $\tau$, it is only changing under limited agent actions and some rare events such as contingencies and maintenance. 

        The agent's actions only influence  $\tt$. Although there are many ways in which  $\tt(t); x(t); y(t)$ could influence (t+1) variables, we only considered the following:

        \begin{itemize}
            \item	$x \to x$: injections are timeseries.
            \item	$\tt \to \tt$ : agent's "position" $\tt(t+1)$ is constrained by past positions $\tt(t)$ given the limited action rule, \textit{de facto} limiting the freedom of the agent to influence y.
            \item	$y \to \tt$: overloaded flows can lead to line disconnections, or cascading failure, hence influencing the topology . 
            \item $\tt \to \aa$: recovery time constrains future actions.
        \end{itemize}

        Without operational constraints and robustness consideration of cascading failures, the problem is mostly an instance of the contextual bandit more specific case on Figure \ref{fig:MDP} (only $x(t) \to x(t+1)$) under which more specific algorithm than RL can be preferred and perform quite well. Adding $\tt \to \tt$ by limiting instantaneous actions makes it a regular RL problem. Our platform could actually run this latter setting as an \textbf{easy mode} to make an agent's training easier at first. However the problem we proposed in this first challenge, our \textbf{hard mode}, already involved some more complexity as depicted on Figure \ref{fig:MDP}. Successful approaches in the easy mode would not necessarily work in the hard mode, especially since it involves robustness issues in the latter case. 
        
        These causal diagrams are hence useful for understanding the problem proposed. It helps the designer anticipate if modeling a new constraint will change the nature and complexity of the problem and it helps the participant select the appropriate class of methods to solve it. In particular, ML and RL methods indeed appeared suited for this first challenge as expected.

    \section{Result Analysis} \label{sec:res}
    
        We now review the challenge results, by first describing the properties of the scenarios on which participants were tested, and giving a description of the best agents. Then we analyze the agents behavior on these and eventually present a post-analysis of their performance. Finally we will propose an oracle approach to derive a near upper-bound for the scores, to better assess the optimization gap of the agents.

        \paragraph{Validation and test scenarios}

            In order to select interesting scenarios, we tried to combine multiple criteria that were identically set for both validation and test scenarios:
            
            \begin{itemize}
                \item Difficulty levels: difficulty ranges from easy (no overload at all for the DN agent) to hard (no known solution has been found using our ensemble of DN$_{\tau}$ agent baseline). It reduces the likeliness of having tie contestants.
                \item Diverse tasks: some scenarios focus on handling overloads in the morning, while others test evening peak consumption management. In a scenario, no overload appears, to test agents does not react randomly. In others, overloads vanish naturally, to test agents do not overreact.
                \item Diverse context: we  include variability by changing the day of the week of the scenarios, to make sure the powergrid can be handled for all  days of the week, and not only at some very restrictive times.
                \item Diverse horizons: we finally have scenarios of different length, varying from 1 day (288 time-steps) to 3 days (864 time-steps), to test agents in different kind of settings: longer scenarios favor stable agents with longer horizon, shorter scenarios favor greedy agents.
            \end{itemize}

            Those scenarios were mainly selected to test the robustness of agents, that is their ability to finish a given scenario. However, beside being robust, participants still had to continuously optimize the grid margins. Let's now describe the best agents and examine their behavior on the test scenarios.

        \paragraph{Agent Description}
        
            From Figure \ref{fig:Leaderboard}, we can see that only the groups called ``Lebron James'' (\textcolor{blue}{LB}), ``LearningRL'' (\textcolor{blue}{LR}) and ``Stephan Curry'' (2nd team from Geirina, so we only consider LB) managed to finish all test scenarios, with ML + RL approaches. Both codes are open-source and referenced on the challenge website \footnote{\label{website}Benchmark competition \& github of winners at \url{l2rpn.chalearn.org}}. ``Kamikaze'' (\textcolor{blue}{KM}, ML approach)  and ``Smart Dispatcher'' (\textcolor{blue}{SD}, Expert system + greedy validation) each failed on one, whereas they previously managed all validation scenarios. ``Menardpr'' (\textcolor{blue}{MD}, greedy tree search approach) failed on one, on both test and validation scenarios. Winners were less computationally intensive at test time, thanks to Machine Learning, while also being more effcicient, making them relevant in a setting of real-time decision making.

            Comparing \textcolor{blue}{LB} and \textcolor{blue}{LR} from a code analysis, \textcolor{blue}{LR} appeared to be the only participant to never query the simulate function to validate or explore additional actions at test time: this is quite an achievement to only rely on what the agent learnt, and trust it. Its model was based on the actor critic (A3C) algorithm \cite{actorcritic}. This is an architecture with two main components, a policy network (actor) and a value network (critic). A3C aim at learning a policy function directly, through the actor module. The critic module then criticize the actor given the new state value after taking an action, to adjust and improve its behavior. Actor and critic modules are learnt asynchronously by pooling multiple workers that learn independently and improve a global agent.

            \textcolor{blue}{LB} agent on the other hand combined an RL model based on a dueling DQN algorithm \cite{duelingDQN}, coupled to a set of actions selected through extended prior analysis and imitation learning. Dueling DQN is most similar to DQN \cite{DBLP:googleDQN} except that the neural network architecture explicitly try to encode separately at its core a state value function and an action advantage function, later combined to better estimate the Q-value.
            
            In addition, \textcolor{blue}{LB} uses few expert rules, especially "don't do anything if all your margins are good enough, below a threshold of 80\%", and make extensive use of the simulate function when deciding on taking an action, to strongly validate a set of suggested action. This is representative on how operators have been doing until today. Their approach is closer to an assistant: an RL model suggests some actions and an expert model make a cautious choice among them and the ones he knows, validating with a simulator.  

            \begin{figure}[!htbp]
                \centering
                \includegraphics[width=9cm, height=4cm]{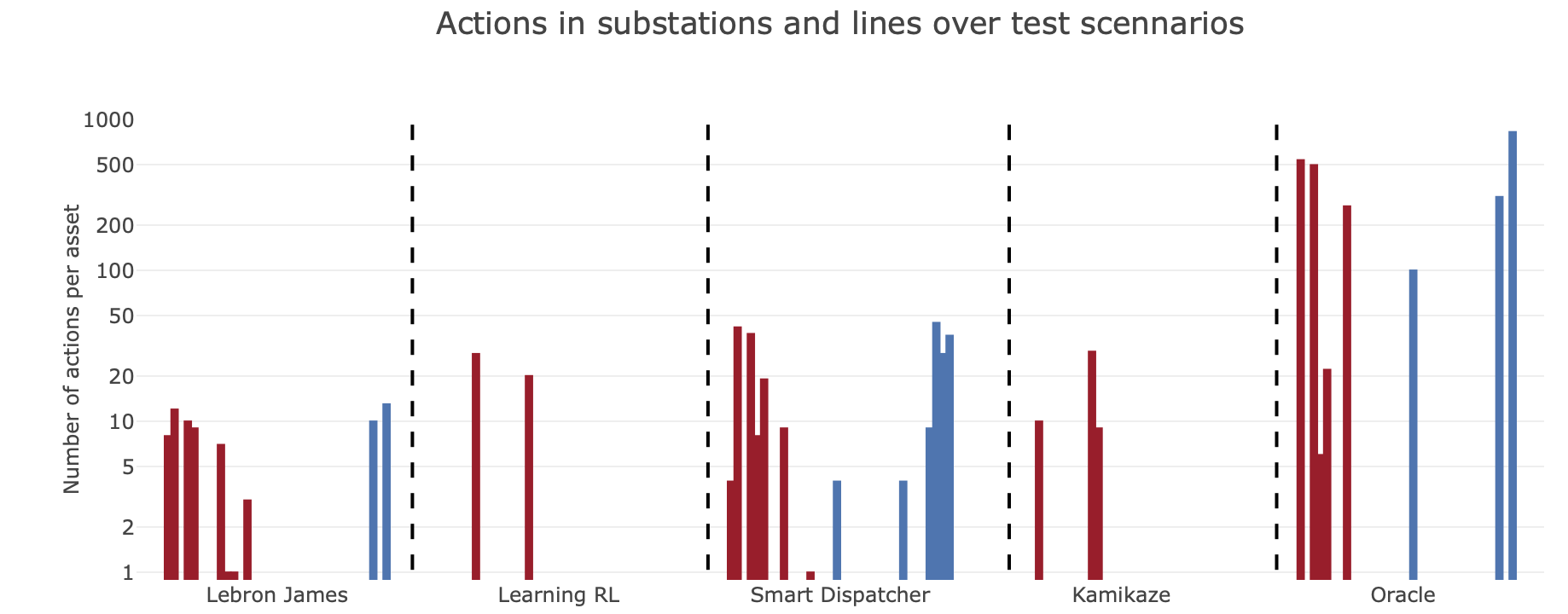}
                
                 \caption{Number of actions used at different substations (red) or lines (blue), ordered by indices, for different agents and our oracle over test scenarios.}
                \label{fig:actions}
            \end{figure}

        \paragraph{Agents action space}

            All participants tried with different strategies to reduce the action space to explore at first. \textcolor{blue}{SD} relied on its operator's expertise to identify relevant substations and lines to test on, still using a diverse set of assets with selected topologies. \textcolor{blue}{MD} only focused on line switching to meet the time constraint with a tree search approach. However, most participants let aside line switching, sometimes an interesting option when no overloads exist to reroute some flows, but often detrimental when the grid is overloaded.

            \textcolor{blue}{LR} also used some domain knowledge over the symmetries in a substation to reduce the number of true actions. More interestingly, they started their exploration from scratch and learnt robust actions automatically through a curriculum learning. They indeed started to learn in an easy mode (no game over), upsampling the scenarios to quickly see more diverse situations. They somehow already learnt useless actions and potentially useful actions to route the flows. They then switched to the hard mode with the appropriate time resolution to learn managing overloads and being robust to them. They used their own reward function when learning to penalize strongly on overloads when occurring, different from our score in (2). Our score is not a good reward function to learn from in that prospect, as its gradient, and hence the learning signal, is null in the overload regime. They eventually converged to only act on two substations, a bit restrictive. This might be changed by relaxing their reward function.

            \textcolor{blue}{LB} team on the other hand did an extensive analysis on influential topologies on a batch of sampled grid states to initialize their learning. At the end, they used quite a diverse set of assets. Figure \ref{fig:actions} summarizes the number and diversity of actions agents used on test scenarios.

        \paragraph{Behavior analysis } 

\begin{figure}
                \centering
                \includegraphics[width=9cm, height=4.5cm]{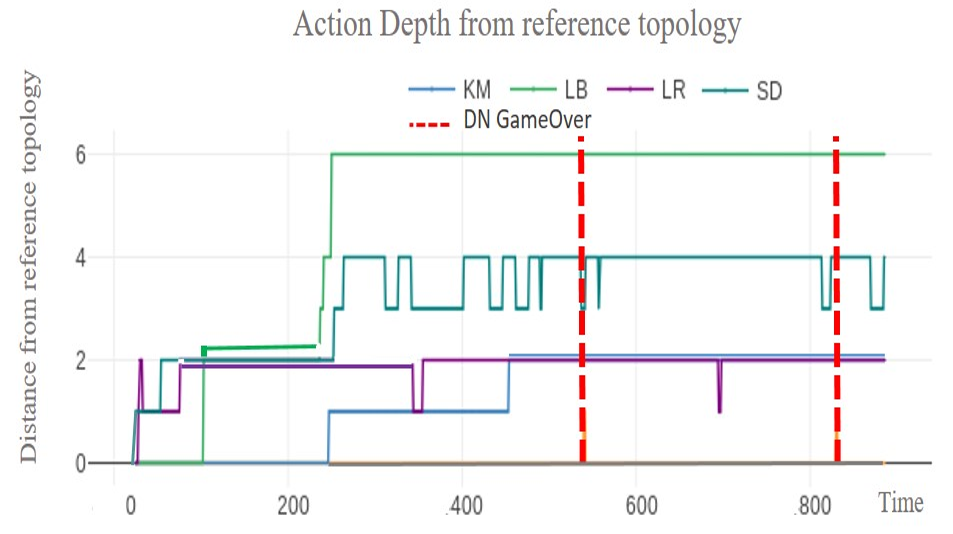}
                \caption{Agent  behavior  over  validation  scenario 3  showing  the  depth  of agent actions at a given time-step, 0 meaning the agent is in the reference topology}
                \label{fig:scenarioBehavior}
            \end{figure}

            Looking at agent actions in real-time from Figure \ref{fig:scenarioBehavior} on a test scenario, we can detect different kinds of behavior. \textcolor{blue}{SD} is indeed doing lots of actions, trying to optimize the score continuously, but somehow going back and forth erratically when the grid is loaded: overloads might be appearing due to its actions. \textcolor{blue}{LB} is a pretty stable agent, anticipating soon enough potential overloads through its expert rule. However its topology seems to always be drifting from the reference topology which might be detrimental on long scenarios. Finally, \textcolor{blue}{LR} is also quite stable due to its small action space but has the ability to go back and forth. Extended AI agent behavior analysis will be conducted in future works, as it is an emerging field \cite{MBehavior} with new tools being released \cite{Bsuite}.

            We eventually ran an additional post analysis experiment on larger batch of monthly scenarios given for training. It showed that all agents are still failing on several scenarios (\textcolor{blue}{LB}: 5 fails, \textcolor{blue}{LR}: 11 fails, \textcolor{blue}{SD}: 9 fails), highlighting some instabilities on the long term. Scenario 37 appeared especially difficult with no agent succeeding. We should hence aposteriori adjust or augment our selection criteria to develop a more comprehensive benchmark of this task. 
            Also, beyond robustness, the challenge was also about continuously optimizing the flows on the grid.



        \paragraph{Revised scores with oracle baseline}

            In the short duration of the competition, participants mainly focused on their agent robustness, as a single gameover prevented them from winning. From the behavior analysis, agents do not appear to take many actions over a scenario, except from \textcolor{blue}{SD} agent, which might indicate that agents did not try to optimize the flows beside avoiding overloads. To assess how good they did on this second task, we need to define an upper-bound baseline to get a better idea of what a good score should be.

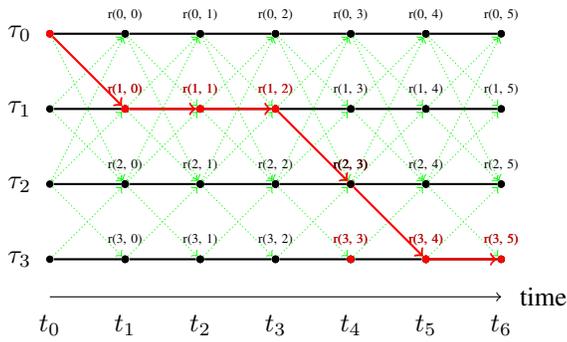
\begin{figure}{}
					\begin{tikzpicture}[dot/.style = {anchor=base,fill,circle,inner sep=1pt}]
	 \tikzstyle{point} = [circle, size=0.01cm, fill=black!80]
    
    \foreach \x in {0,1,2,3,4,5} 
    { 
        
        \foreach \y in {0,1,2,3}
        {
                \node[dot] (\x \y) at (\x,\y){};
                \node[dot] (\x o \y) at (1+\x,\y){};
                \node[dot,label={above:{\tiny r(\y, \x)}}] (\x aa \y) at (1+\x,3-\y){};
                \draw[thick] (\x \y) -- (\x o \y);
        }
    }
    
    \foreach \x in {0,1,2,3,4,5,6}{
         \node[label={below:$t_\x$}] (\x axis) at (\x,-0.5){};
    }
    
    \node[label={right:time}] (6 axis) at (6,-0.5){};

    \draw[->, color=black]
                (0, -0.5) -- (6,-0.5);
    
    \foreach \y in {0,1,2,3}
    {   \node[label=left:$\tau_\y$](0e\y) at (0, 3-\y){};  }

    {\foreach \x in {0,1,2,3,4,5} 
    { 
    	\foreach \y/\z in {0/1, 1/2, 1/3, 2/3}
        	{   
        	\draw[densely dotted, ->, color=green]
                (\x \y) -- (\x o \z);
            \draw[densely dotted,->, color=green]
                (\x \z) -- (\x o \y);
    	}
	}}

	\foreach \x/\y/\z/\newY in{ 0/3/2/1}
		{
		{\draw[->, thick, color=red]
                (\x \y) -- (\x o \z);}
        \node[dot,color=red] (\x \y) at (\x,\y){};
        \node[dot,color=red] (\x aa \y) at (\x,\y){};
        \node[dot,color=red,label={above:{\textcolor{red}{ \tiny r(\newY, \x)}}}] (\x aa \y) at (\x+1,\z){};

		}
	
	\foreach \x/\y/\z/\newY in { 1/2/2/1, 2/2/2/1, 3/2/1/2, 4/1/0/3, 5/0/0/3}
		{
		{\draw[->, thick, color=red]
                (\x \y) -- (\x o \z);}
        \newcount\cnt
        \cnt=\numexpr1+1\relax
        
        \node[dot,color=red] (\x \y) at (\x,\y){};
        \node[dot,color=red] (\x aa \y) at (\x,\y){};
        \node[dot,color=red,label={above:{\textcolor{red}{ \tiny r(\newY, \x)}}}] (\x aa \y) at (\x+1,\z){};

		}
	
	\foreach \x/\y/\z/\newY in {4/1/0/2}
		{
		
        
        \node[dot,label={above:{ \tiny r(2, 3)}}] (\x aa \y) at (\x,\y){};
        \node[dot,color=red,label={above:{\textcolor{red}{ \tiny r(3, 3)}}}] (\x aa \y) at (\x,\y-1){};

		}

    \end{tikzpicture}
    \caption{reward graph $\mathcal{G}$ for several topology configurations $\tau$ and allowed transitions in dashed green. The oracle optimal path (in red) misses a max immediate reward at $t_3$ as no direct transition is allowed between $\tau_1$ and $\tau_3$.}
       \label{fig:OracleGraph}
	\end{figure}

            While imperfect information is given to the participants along the game as they don't  know yet the future, as organizers we can make use of the full scenario information to compute an oracle baseline, our upper-bound. For our method, we rely on the connection there exists between topology configuration and topological action: given one action we can deduce the topology configuration we reach, knowing which configuration to reach to get some reward we can deduce the necessary action. We decided to run test scenarios under thousands of topology configurations in parallel (not all, which is too hard computationally) to later identify what would have been very good topologies at a given time and infer the preferred course of actions. Ultimately, it is framed as a longest path problem as on Figure \ref{fig:OracleGraph}. Our method can be decomposed in 5 steps:
            
            \begin{enumerate}
                \item Define a dictionary of interesting unitary actions $D_a$ from $n$ unitary selected assets as on Table \ref{table:tables}
                \item Identify all combinations of $D_a$ actions to create the oracle action space $A_{oracle}$ and related topology configuration space $S(\tau)_{oracle}$ when applied to $\tau_{ref}$. A topology can be $n$ action away from the reference one.
                \item Apply all $\tau \in  S(\tau)_{oracle}$ independently and run them in parallel on scenarios to compute the reward of each configuration $\tau$ at each time-step t. This results in directed chains $\mathcal{G_\tau}$ with edges $e_{G_{\tau}}(t)=reward(\tau,t)$
                \item From all \{$\mathcal{G_\tau}$\}, build the overall connex graph $\mathcal{G}$ of possible topology trajectories, given allowed topology transitions from operational rule. Add edges between reachable configurations ($\tau_a$,$\tau_b$): $e_{(\tau_a,\tau_b)}(t)=reward(\tau_b,t)$.
                \item Compute the best score through the longest path on the directed acyclic graph $\mathcal{G}$ and determine the related course of actions $\{a(t)\}_{t=0..T} \in A_{oracle}$.
            \end{enumerate}

Our score is always greater than all agents while respecting the operational rules, effectively defining on upper-bound. To determine how good other agents did, we define a revised score taking DN as a zero-reference score (computed in easy mode considering the optimization task only), oracle as a max:
\begin{equation}
Score_{Normalized} =\frac{Score_{agent}-Score_{DN}}{Score_{oracle}-Score_{DN}}
\end{equation}
The revised scores in Table \ref{table:tables} suggest that there still exists a gap between our oracle upper-bound and the best submitted agents. Figure \ref{fig:actions} highlights that our oracle continuously took some actions, every 2 to 3 timesteps on average, as opposed to other agents. This confirms that agents did not do quite well on this optimizing task.  Thus we reopened the challenge test case as a benchmark\footref{website}. This highlights that controlling the topology still remains a hard problem given a huge action space requiring lots of exploration.
It also challenges us as organizers to offer more representative scores that will give more incentives to the participants to perform better on a related task.

\begin{table}

\hfill%
\subfloat[Selected unitary actions at subs and lines for oracle. In bold, the ones used previsously for thermal limit design]{
\resizebox{0.47\linewidth}{!}{
     \begin{tabular}{ | l | c | l | }\toprule
     \textbf{Felxibility} & 
     \textbf{id} &
     \textbf{Target config.} \\ \midrule
     \multirow{2}{*}{Sub 6} 
                           & \textbf{1} & \textbf{[0, 0, 0, 0, 1, 1]} \\
                           & 2 & [0, 1, 0, 0, 1, 1] \\ \midrule
     \multirow{2}{*}{Sub 5} 
                           & \textbf{1} & \textbf{[0, 1, 0, 0, 1]} \\
                           & 2 & [0, 0, 1, 0, 1] \\ \midrule
     \multirow{4}{*}{Sub 4} 
                           & \textbf{1} & \textbf{[0, 0, 1, 0, 1, 0]} \\
                           & 2 & [0, 0, 0, 1, 1, 0] \\
                           & \textbf{3} & \textbf{[1, 0, 1, 0, 1, 1]} \\
                           & 4 & [1, 0, 1, 0, 1, 0] \\ \midrule
     \multirow{3}{*}{Sub 9} 
                           & \textbf{1} & \textbf{[1, 0, 1, 0, 1]} \\
                           & 2 & [0, 0, 1, 0, 1] \\
                           & 3 & [1, 1, 0, 1, 1] \\ \midrule
     \multirow{2}{*}{Sub 2} 
                           & 1 & [1, 1, 0, 1, 0, 1] \\
                           & 2 & [1, 1, 0, 1, 0, 0] \\ \midrule
     \multirow{1}{*}{\shortstack{Lines 2-4, 5-6\\ 10-11, 13-14}} & 0 & [0] \\
                             & 1 & [1]  \\
                             \bottomrule
     \end{tabular}
}
}
\subfloat[Normalized scores for SD and LB agents, compared to an oracle with 100 points. There still exists an optimization gap to improve on. ]{
\resizebox{0.47\linewidth}{!}{
     \begin{tabular}{ | l | c  c  |}
     \hline
     \textbf{Scenario} & \textbf{SD} & \textbf{LB} 
     \\ \hline
     \textbf{1} & 72,5 & 61,5  \\ \hline
     \textbf{2} & -10,5 & 90,2  \\ \hline 
     \textbf{3} & 53 & 82,5  \\ \hline
     \textbf{4} & 49,5 & 81,5  \\ \hline
     \textbf{5} & 47,5 & 70,0  \\ \hline
     \textbf{6} & 48 & 47  \\ \hline
     \textbf{7} & 19,5 & 63  \\ \hline
     \textbf{8} & 39,5 & 77,5  \\ \hline
     \textbf{9} & 52,5 & 93  \\ \hline
     \textbf{10} & 56,5 & 56,5  \\ \hline
     \end{tabular}
}
}
\caption{}
\label{table:tables}
\end{table}

\section{Conclusions}
The challenge was successful in addressing safety considerations and was necessary to open a new research avenue for a broad community, extended to Machine Learning researchers.
It demonstrated that developing topological controllers for real-time decision making is indeed possible, especially when using reinforcement learning. Framing the problem as a two-factor MDP allowed us to also expose the difficulties faced by reinforcement learning solutions to such control problems. The diversity of submissions and behaviors helped us appreciate the {\em pros} and {\em cons} of each approach. Evaluating the participants' performance pushed us to define new interesting baselines and scoring metrics for future research and challenge designs. Our post-challenge analyses revealed both the feasibility of such approaches and the important gap to optimality, particularly for {\em continuous} power flow optimization, giving us an incentive to take the design of a new benchmark to the next level, including scaling up the dimensions of the grid. 

\subsection*{Acknowledgements}
{\footnotesize Many people have contributed to the design and implementation of the L2RPN challenge. We would like to acknowledge the contributions of Marc Mozgawa, Kimang Khun, Joao Araùjo, Marc Schoenauer, Patrick Panciatici,
Olivier Pietquin, and Gabriel Dulac-Arnold.
The challenge is running on the Codalab platform, administered by Université Paris-Saclay and maintained by CKCollab LLC, with primary developers Eric Carmichael and
Tyler Thomas}

\bibliographystyle{abbrv}
\bibliography{references.bib}

\end{document}